\documentclass[twocolumn,english,american,aps,pra,10pt,floatfix]{revtex4-2}
\usepackage[T1]{fontenc}
\usepackage[utf8]{inputenc}
\usepackage{babel}
\usepackage{amsmath}
\usepackage{amssymb}
\usepackage{graphicx}
\usepackage[pdfusetitle,
 bookmarks=true,bookmarksnumbered=false,bookmarksopen=false,
 breaklinks=false,pdfborder={0 0 1},backref=false,colorlinks=false]
 {hyperref}

\makeatletter

\providecommand{\tabularnewline}{\\}

\usepackage{tikz}
\usetikzlibrary{quantikz2}
\usepackage{diagbox}
\usepackage{orcidlink}
\hypersetup{hidelinks}
\makeatother

\begin{document}
\selectlanguage{english}%
\global\long\def\ket#1{|#1\rangle}%

\global\long\def\bra#1{\langle#1|}%

\global\long\def\braket#1#2{\langle#1|#2\rangle}%

\global\long\def\tr#1{\text{tr}#1}%

\title{Probabilistic imaginary-time evolution via compiled diagonal representations}
\author{Hengyue Li \orcidlink{0009-0002-6903-2994}}
\email{lihy@arclightquantum.com}

\affiliation{Arclight Quantum Computing Inc., Beijing, 100191, People's Republic
of China}
\author{Huazhe Lou}
\affiliation{Arclight Quantum Computing Inc., Beijing, 100191, People's Republic
of China}
\author{Yang Li}
\affiliation{Arclight Quantum Computing Inc., Beijing, 100191, People's Republic
of China}
\author{Mingyu Huang}
\affiliation{Arclight Quantum Computing Inc., Beijing, 100191, People's Republic
of China}
\author{Yusheng Yang\orcidlink{0000-0002-3584-4954}}
\affiliation{Arclight Quantum Computing Inc., Beijing, 100191, People's Republic
of China}
\author{Jianyong Sheng}
\affiliation{Arclight Quantum Computing Inc., Beijing, 100191, People's Republic
of China}
\author{Jian Sun}
\affiliation{Arclight Quantum Computing Inc., Beijing, 100191, People's Republic
of China}
\author{Shenggang Ying}
\email{yingsg@ios.ac.cn}

\affiliation{Institute of Software, Chinese Academy of Sciences, Beijing, 100190,
People's Republic of China}
\affiliation{Arclight Quantum Computing Inc., Beijing, 100191, People's Republic
of China}
\begin{abstract}
We introduce a probabilistic imaginary-time evolution scheme based
on a compiled approximate diagonal representation of the Hamiltonian.
A basis transformation is optimized offline. For a fixed compiled
representation, the online circuit structure and gate count per attempt
are independent of the imaginary time, and the same representation
can be reused across different input states and evolution times. We
derive a condition for enhanced success probability. Numerical tests
show reduced online cost relative to termwise Pauli-based methods
at a common accuracy threshold. We also examine how approximation
errors and offline preparation costs affect the overall benefit.
\end{abstract}
\maketitle

\section{Introduction}\label{sec:introduction}

Imaginary-time evolution (ITE) is a tool for state preparation and
equilibrium calculations in quantum many-body systems \citep{Motta2020}.
For a Hamiltonian $H$, the operator $e^{-\beta H}$, with imaginary
time $\beta\geq0$, exponentially suppresses higher-energy components
relative to lower-energy ones. This allows ground-state preparation
from inputs with nonzero ground-state overlap. Imaginary-time evolution
also serves as a subroutine for thermal-state preparation and the
calculation of finite-temperature observables \citep{White2009,Getelina2023}.
Because this evolution is nonunitary and generally changes the state
norm, it cannot be implemented directly by a deterministic unitary
circuit. The task is therefore to realize the corresponding normalized
evolution with quantum circuits.

Several quantum algorithms have been developed to implement ITE. Quantum
imaginary-time evolution (QITE) approximates short steps of the normalized
evolution using state-dependent unitary updates constructed from measurements
\citep{Motta2020}. Variational ITE (VITE) approximates the trajectory
within a parameterized circuit ansatz, whose parameters are updated
according to a variational principle \citep{McArdle2019}. Probabilistic
ITE (PITE) implements nonunitary imaginary-time filters through joint
unitary evolution of the system and ancillary qubits, followed by
postselection on the ancillary measurement outcomes \citep{Kosugi2022,Xie2024,Leadbeater2024}.
The present work builds on the PITE framework. For a given Hamiltonian
and evolution time, the same filter circuit can be applied to different
input states. Its practical cost depends on both the circuit depth
and the probability of successful postselection.

For noncommuting Hamiltonians, termwise PITE approximates the imaginary-time
propagator by repeated products of filters for individual Pauli terms
\citep{Xie2024,Leadbeater2024}. Reducing the time step controls the
product-formula error but increases the circuit depth needed to reach
a given total imaginary time. The normalization required for individual
filters also affects the probability of successful postselection and
hence the number of repetitions. A diagonal representation can simplify
the circuit because its terms commute, allowing repeated time steps
to be combined into one filter per term \citep{Kim2026FinITE}. A
suitable change of basis may also reduce the normalization loss and
improve the success probability. For Hamiltonians that are not diagonal
in the computational basis, realizing these potential benefits requires
an accurate, compact diagonal representation and a shallow circuit
for the basis transformation.

Here we introduce fast ITE (FITE), a probabilistic scheme based on
an approximate diagonal representation of the Hamiltonian. We optimize
the basis transformation offline and determine the diagonal coefficients
from the physical Hamiltonian. For a fixed compiled representation,
the circuit can be reused across different input states and imaginary
times by adjusting the filter angles, with its depth and gate count
per attempt independent of the total imaginary time $\beta$. We derive
the condition for enhanced success probability relative to termwise
Pauli filtering of the original Hamiltonian. Numerical tests on small
systems show a lower online gate cost per accepted output than these
termwise baselines at a common accuracy threshold, including the cost
of failed attempts. We also examine how approximation errors and offline
preparation costs affect the practical benefit.

We first describe the method, then compare its accuracy, success probability
and online cost with the baselines, and discuss the conditions for
practical benefit. Circuit details, derivations and additional numerical
checks are collected in the appendices.

\section{Method}\label{sec:method}

We consider an $n$-qubit Hamiltonian $H=\sum_{j=1}^{N_{P}}c_{j}P_{j}$,
where the $P_{j}$ are distinct nonidentity Pauli strings and the
$c_{j}$ are real coefficients with energy units. The identity component
is omitted because a constant energy shift does not affect normalized
ITE. We use $\|\cdot\|_{2}$ for the operator norm of matrices and
the Euclidean norm of state vectors, and $\|\cdot\|_{1}$ for the
coefficient-vector one-norm. For a normalized input $\ket{\psi}$,
our target is

\begin{equation}
\ket{\psi_{\beta}}=\frac{e^{-\beta H}\ket{\psi}}{\|e^{-\beta H}\ket{\psi}\|_{2}}.\label{eq:target-ite-state}
\end{equation}

We set $\hbar=1$ and use $\beta\geq0$ for the total imaginary time.
The method separates offline preparation of a diagonal representation
from online filtering. We first describe the filter construction and
then explain how the basis transformation and diagonal coefficients
are obtained.

\subsection{Imaginary-time evolution in a diagonal basis}\label{sec:diagonal-filters}

Suppose that a unitary basis transformation $W$ yields the approximate
representation

\begin{equation}
H\simeq\widetilde{H}=W\mathcal{H}W^{\dagger}.\label{eq:diagonal-model}
\end{equation}

We write $\mathcal{H}=\sum_{k=1}^{K}\xi_{k}\mathcal{Z}_{k}$, where
the $\xi_{k}$ are real coefficients and the $\mathcal{Z}_{k}$ are
$K$ retained, distinct nonidentity Pauli-$Z$ strings. Explicitly,

\begin{equation}
\mathcal{Z}_{k}=\bigotimes_{q=1}^{n}Z^{s_{kq}},\qquad s_{kq}\in\{0,1\},\label{eq:diagonal-pauli-string}
\end{equation}

where $Z^{0}=I$ and $Z^{1}=Z$, so $s_{kq}$ specifies the action
on qubit $q$. At least one $s_{kq}$ is nonzero for each retained
string. Since these operators commute, the corresponding imaginary-time
propagator factorizes as

\begin{equation}
e^{-\beta\widetilde{H}}=W\left(\prod_{k=1}^{K}e^{-\beta\xi_{k}\mathcal{Z}_{k}}\right)W^{\dagger}.\label{eq:diagonal-propagator}
\end{equation}

This factorization is exact for the represented Hamiltonian $\widetilde{H}$
and introduces no time-discretization error. Each factor covers the
full imaginary time $\beta$, so one filter per retained term suffices.

We implement each factor probabilistically with an ancilla-assisted
Pauli filter \citep{Xie2024,Leadbeater2024}. Postselecting the ancilla
outcome zero gives the Kraus operator

\begin{equation}
\mathcal{K}_{k}=e^{-\beta|\xi_{k}|}e^{-\beta\xi_{k}\mathcal{Z}_{k}}.\label{eq:diagonal-kraus}
\end{equation}

The scalar prefactor normalizes the operator norm to unity. Define
the coefficient norms $\lambda=\sum_{j=1}^{N_{P}}|c_{j}|$ and $\Lambda=\sum_{k=1}^{K}|\xi_{k}|$
for the physical and diagonal representations, respectively. The successful
branch of the complete circuit is

\begin{equation}
\mathcal{K}_{{\rm FITE}}=W\left(\prod_{k=1}^{K}\mathcal{K}_{k}\right)W^{\dagger}=e^{-\beta\Lambda}e^{-\beta\widetilde{H}}.\label{eq:fite-construction}
\end{equation}

Conditioned on all filters succeeding, the normalized output is $\ket{\widetilde{\psi}_{\beta}}=e^{-\beta\widetilde{H}}\ket{\psi}/\|e^{-\beta\widetilde{H}}\ket{\psi}\|_{2}$.
The probability of this event is

\begin{equation}
P_{{\rm FITE}}(\psi)=e^{-2\beta\Lambda}\bra{\psi}e^{-2\beta\widetilde{H}}\ket{\psi}.\label{eq:fite-success-probability}
\end{equation}

\begin{figure*}
\centering
\centering{}\includegraphics[width=1\textwidth]{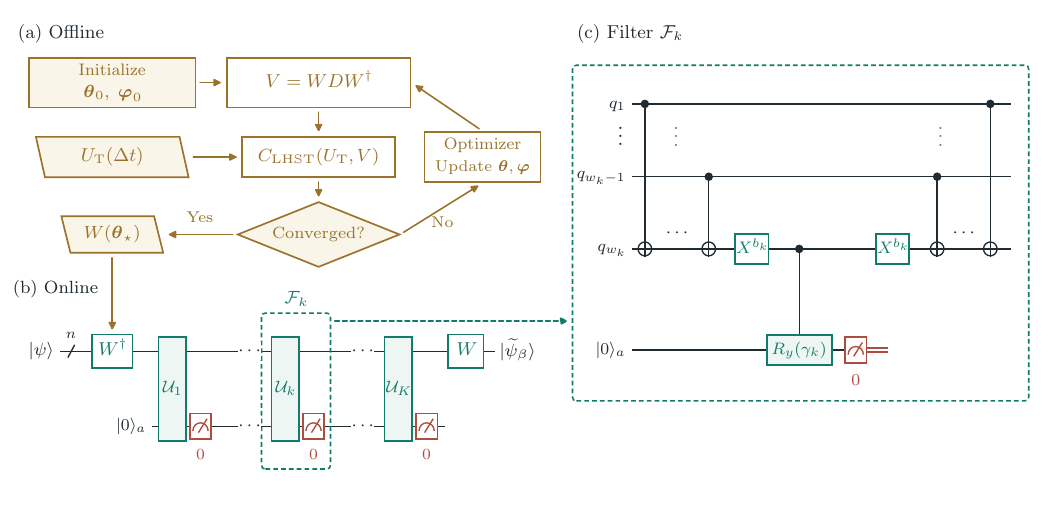}\caption{Overview of FITE. (a) Offline compilation starts from $\boldsymbol{\theta}_{0},\boldsymbol{\varphi}_{0}$
and minimizes $C_{{\rm LHST}}(U_{{\rm T}},V)$ for $V=WDW^{\dagger}$.
The decision tests the optimizer stopping criterion; $W(\boldsymbol{\theta}_{\star})$
is the retained basis circuit. (b) Each filter block $\mathcal{F}_{k}$
comprises a joint unitary $\mathcal{U}_{k}$ and ancilla postselection,
with $_{a}\langle0|\mathcal{U}_{k}|0\rangle_{a}=\mathcal{K}_{k}$.
All $K$ outcomes must be zero; otherwise the attempt is aborted and
the same input is prepared again. The ancilla is reused; reset operations
are omitted from the drawing, since a successful readout leaves it
in $\protect\ket 0$. (c) For $\mathcal{Z}_{k}=\prod_{j=1}^{w_{k}}Z_{q_{j}}$,
CX gates encode parity onto $q_{w_{k}}$. Here $w_{k}$ is the number
of nonidentity factors; unaffected qubits are omitted. For nonzero
$\xi_{k}$, $b_{k}=[1+\operatorname{sgn}(\xi_{k})]/2$, and the controlled
rotation uses $\gamma_{k}$ from Eq. (\ref{eq:sm-filter-angle}).
The second $X^{b_{k}}$ and reversed CX network restore the system.
Ancilla measurement commutes with this restoration and is shown first
to permit early abort. A zero coefficient gives the identity filter;
for $w_{k}=1$, the CX networks are absent.}\label{fig:fite-workflow}
\end{figure*}

Figure \ref{fig:fite-workflow}(b) shows the complete online circuit,
and panel (c) expands a general diagonal filter. The $K$ filters
share a single ancilla, which is measured after each filter and reset
before the next. For fixed $W$ and retained diagonal terms, varying
$\beta$ changes only the filter angles, leaving the circuit depth
and gate count per attempt unchanged. The remaining task is to determine
$W$ and the coefficients $\{\xi_{k}\}$.

\subsection{Variational compilation of real-time evolution}\label{sec:basis-compilation}

Several approaches can be used to construct a diagonal representation
of $H$, including variational Hamiltonian diagonalization \citep{Commeau2020VHD}
and Cartan decomposition for structured Hamiltonians \citep{Kokcu2022Cartan}.
Here we use variational fast forwarding (VFF) \citep{Crstoiu2020}
as a simple implementation for obtaining the basis transformation
$W$.

The Hamiltonian $H$ and its exact real-time propagator $e^{-i\Delta tH}$
share an eigenbasis, allowing the basis transformation to be learned
through a unitary compilation problem. For a short interval $\Delta t>0$,
let $U_{{\rm T}}(\Delta t)$ denote a circuit approximation to this
propagator, such as a product-formula step. VFF fits the factorization

\begin{equation}
U_{{\rm T}}(\Delta t)\simeq V(\boldsymbol{\theta},\boldsymbol{\varphi})=W(\boldsymbol{\theta})D(\boldsymbol{\varphi})W^{\dagger}(\boldsymbol{\theta}),\label{eq:variational-compilation}
\end{equation}

Here $D(\boldsymbol{\varphi})=\prod_{k=1}^{K}e^{-i\varphi_{k}\mathcal{Z}_{k}/2}$
uses the same retained diagonal strings as the filters. We optimize
the basis parameters $\boldsymbol{\theta}$ and diagonal angles $\boldsymbol{\varphi}$
by minimizing the local Hilbert--Schmidt test cost $C_{{\rm LHST}}$.
This cost measures the mismatch between the two unitaries and vanishes
when they agree up to a global phase \citep{Khatri2019}. The optimization
supplies the basis circuit $W=W(\boldsymbol{\theta}_{\star})$, where
$\boldsymbol{\theta}_{\star}$ denotes the retained optimized basis
parameters.

The optimized angles do not uniquely specify the energy coefficients
needed for imaginary-time filtering. Shifting any $\varphi_{k}$ by
$2\pi$ changes $D$ only by a global sign, leaving the compilation
cost unchanged. Even for an exact termwise match to $e^{-i\Delta tH}$,
the wrapped angle $\varphi_{k}^{{\rm wrap}}=\operatorname{Arg}(e^{i\varphi_{k}})$
admits the energy branches

\begin{equation}
\xi_{k}^{(m_{k})}=\frac{\varphi_{k}^{{\rm wrap}}+2\pi m_{k}}{2\Delta t},\qquad m_{k}\in\mathbb{Z}.\label{eq:energy-branches}
\end{equation}

In addition, errors in the target $U_{{\rm T}}$ and its variational
fit can affect the inferred coefficients even when a phase branch
is selected. The optimized phases therefore do not in general yield
accurate physical coefficients $\xi_{k}$. We instead retain the optimized
basis $W$ and determine its retained diagonal coefficients directly
from $H$.

\subsection{Recovering the diagonal Hamiltonian coefficients}\label{sec:coefficient-recovery}

With the compiled basis $W$ fixed, we determine the diagonal coefficients
using Pauli orthogonality. Let $N=2^{n}$. The retained strings satisfy
$\operatorname{Tr}[\mathcal{Z}_{\ell}\mathcal{Z}_{k}]=N\delta_{\ell k}$,
so multiplying the expansion of $\mathcal{H}$ by $\mathcal{Z}_{k}$
and taking the normalized trace isolates its coefficient:

\begin{equation}
\frac{1}{N}\operatorname{Tr}[\mathcal{H}\mathcal{Z}_{k}]=\sum_{\ell=1}^{K}\xi_{\ell}\delta_{\ell k}=\xi_{k}.\label{eq:coefficient-extraction}
\end{equation}

Applying this projection directly to the physical Hamiltonian in the
compiled basis defines

\begin{equation}
\xi_{k}^{{\rm proj}}=\frac{1}{N}\operatorname{Tr}[W^{\dagger}HW\mathcal{Z}_{k}].\label{eq:projected-coefficients}
\end{equation}

For an exact diagonal representation within the retained support,
this projection recovers the diagonal coefficients exactly. For an
approximate representation, it extracts the retained diagonal Pauli
components of $W^{\dagger}HW$. The remaining representation error
comes from off-diagonal components and omitted diagonal terms.

Substituting the Pauli expansion $H=\sum_{j}c_{j}P_{j}$ and using
trace cyclicity expresses each coefficient as a weighted sum of normalized
traces:

\begin{equation}
\xi_{k}^{{\rm proj}}=\sum_{j}c_{j}\frac{1}{N}\operatorname{Tr}[\mathcal{Z}_{k}W^{\dagger}P_{j}W].\label{eq:coefficient-traces}
\end{equation}

\begin{figure}
\centering
\centering{}\includegraphics[width=1\columnwidth]{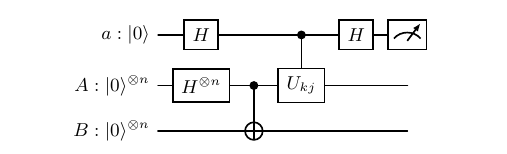}\caption{The circuit estimates the normalized trace of $U_{kj}=\mathcal{Z}_{k}W^{\dagger}P_{j}W$
using $2n+1$ qubits. Registers $A$ and $B$ are prepared in $n$
Bell pairs, and the ancilla readout satisfies Eq. (\ref{eq:normalized-trace-readout}).
The controlled $U_{kj}$ can be implemented with unconditional $W$
and $W^{\dagger}$ together with controlled Pauli gates.}\label{fig:coefficient-trace-test}
\end{figure}

Each operator $U_{kj}=\mathcal{Z}_{k}W^{\dagger}P_{j}W$ is unitary,
allowing its normalized trace to be estimated with the test in Fig.
\ref{fig:coefficient-trace-test} \citep{Knill1998}. Registers $A$
and $B$ are prepared in $n$ Bell pairs, giving the reduced state
$\rho_{A}=I/N$. The Hadamard test on the ancilla then yields

\begin{equation}
p_{0}^{(kj)}-p_{1}^{(kj)}=\frac{1}{N}\operatorname{Re}\operatorname{Tr}U_{kj},\label{eq:normalized-trace-readout}
\end{equation}

where $p_{0}^{(kj)}$ and $p_{1}^{(kj)}$ are the probabilities of
its two measurement outcomes. The trace is real because $\mathcal{Z}_{k}$
and $W^{\dagger}P_{j}W$ are Hermitian. Combining this readout with
Eq. (\ref{eq:coefficient-traces}) gives

\begin{equation}
\xi_{k}^{{\rm proj}}=\sum_{j}c_{j}\left[p_{0}^{(kj)}-p_{1}^{(kj)}\right].\label{eq:coefficient-readout}
\end{equation}

The resulting coefficients are used as $\xi_{k}=\xi_{k}^{{\rm proj}}$
in the diagonal filters. This procedure uses the physical Hamiltonian
directly and avoids selecting energy branches from the optimized phases.
Finite-shot coefficient estimation and its sampling cost are discussed
in Appendix \ref{app:coefficient-estimation}.

\subsection{Workflow summary}\label{sec:workflow-summary}
\begin{enumerate}
\item \textbf{Compile the basis.} For a given Hamiltonian $H$, we first
choose the basis ansatz $W(\boldsymbol{\theta})$ and retained diagonal
strings $\{\mathcal{Z}_{k}\}$. Short-time unitary compilation then
determines $W$.
\item \textbf{Recover the physical coefficients.} Keeping $W$ fixed, we
recover the coefficients $\xi_{k}=\xi_{k}^{\mathrm{proj}}$ from the
physical Hamiltonian using the trace test and weighted readout in
Eq. (\ref{eq:coefficient-readout}). Basis compilation and coefficient
recovery together constitute the offline preparation.
\item \textbf{Construct the filters.} For the requested imaginary time $\beta$,
the recovered coefficients specify the commuting filters, one for
each retained $\mathcal{Z}_{k}$. Each filter covers the full $\beta$,
so the sequence contains $K$ filters.
\item \textbf{Execute and reuse.} The online circuit executes this sequence
with one ancilla measured and reset between filters. All-zero outcomes
yield the normalized output $\ket{\widetilde{\psi}_{\beta}}$; failed
attempts are discarded. For fixed $H$, the compiled $W$ and recovered
coefficients can be reused across input states and imaginary times.
Changing $\beta$ only changes the filter angles, leaving the circuit
structure per attempt unchanged.
\end{enumerate}

\section{Numerical results}\label{sec:numerics}

\subsection{Models and protocol}\label{sec:protocol}

We consider a two-site Hubbard model and open XY and XYZ chains, following
the model families studied in VFF \citep{Crstoiu2020}. The Hubbard
model provides a compact example of interacting fermions; the XY chain
allows an analytic check of the normalization mechanism; the XYZ chain
adds anisotropic exchange, $ZZ$ interactions and a longitudinal field.
The Hamiltonians are

\begin{align}
H_{{\rm Hub}} & =-X_{1}-X_{2}+3Z_{1}Z_{2},\nonumber \\
H_{{\rm XY}} & =-\sum_{j=1}^{n-1}(X_{j}X_{j+1}+Y_{j}Y_{j+1}),\nonumber \\
H_{{\rm XYZ}} & =\sum_{j=1}^{n-1}(X_{j}X_{j+1}+1.1Y_{j}Y_{j+1}\nonumber \\
 & \qquad+1.2Z_{j}Z_{j+1})+1.3\sum_{j=1}^{n}Z_{j}.\label{eq:models}
\end{align}

Adapting the ansatz to Hamiltonian structure, including its symmetries,
can improve variational compilation \citep{Crstoiu2020}. Since systematic
ansatz selection remains an open issue outside our present focus,
we use a common default layered ansatz family. Appendix \ref{app:circuits}
specifies the circuits and optimization settings. We encode the Hubbard
model in two qubits and use the main configurations $(n,d,K)=(2,2,3)$,
$(5,6,5)$ and $(3,8,6)$ for Hubbard, XY and XYZ, respectively. Here,
$n$ is the number of system qubits, $d$ is the number of layers
in the basis circuit and $K$ is the number of retained diagonal terms.
The compilation target is a first-order Trotter step $U_{{\rm T}}(\Delta t)$
with $\Delta t=0.05$.

Physical-Hamiltonian projection determines the diagonal coefficients
directly. For the main numerical comparisons, we use the phase-conversion
shortcut $\xi_{k}^{\mathrm{phase}}=\varphi_{k}^{\mathrm{wrap}}/(2\Delta t)$.
Unless stated otherwise, the numerical results use $\xi_{k}=\xi_{k}^{\mathrm{phase}}$;
$\widetilde{H}$ and $\Lambda$ refer to the same coefficient set.
With $E_{b}=\max\{\lambda,\Lambda\}$, using the coefficient norms
defined in Sec. \ref{sec:diagonal-filters}, the main Hubbard, XY
and XYZ fits give $\Delta tE_{b}=0.250$, 0.400 and 0.525, respectively,
all below $\pi/2$.

We compare FITE with first-order Pauli PITE (P1), symmetric second-order
Pauli PITE (P2), and symmetric second-order filtering of nearest-neighbor
two-site blocks (G2) \citep{Xie2024,Leadbeater2024}. P1/P2 test the
effect of product-formula order, while G2 tests local grouping and
its normalization. Their imaginary-time step $\tau=\beta/M$, where
$M$ is the number of steps, is independent of the real-time compilation
interval $\Delta t$. For the accuracy and success-probability comparison,
we fix $\tau=0.05$; for the resource comparison, we select $M$ at
a common accuracy threshold. Schedules and normalizations are given
in Appendix \ref{app:resources}.

We assess both the accuracy of the accepted state and its success
probability. For the Kraus operator $\mathcal{K}$ associated with
successful postselection, we define $P(\psi)=\|\mathcal{K}\ket{\psi}\|_{2}^{2}$
and $\ket{\psi_{\beta}^{{\rm out}}}=\mathcal{K}\ket{\psi}/\sqrt{P(\psi)}$.
Fidelity is measured against exact ITE of the same input, $F(\psi)=|\langle\psi_{\beta}|\psi_{\beta}^{{\rm out}}\rangle|^{2}$;
the FITE output is $\ket{\widetilde{\psi}_{\beta}}$. Unless stated
otherwise, bars denote complex-Haar input averages: $\overline{F}=\mathbb{E}_{\psi}F(\psi)$
and $\overline{P}=2^{-n}\operatorname{Tr}(\mathcal{K}^{\dagger}\mathcal{K})$.
We estimate mean infidelity using the same 16,384 inputs per model
for all methods, with error bands of $\pm1.96$ standard errors where
shown, and evaluate mean success exactly. The main benchmarks use
noiseless circuits and exclude sampling noise in coefficient estimation.

\subsection{Accuracy and success probability}\label{sec:accuracy}

\begin{figure}
\centering
\centering{}\includegraphics[width=1\columnwidth]{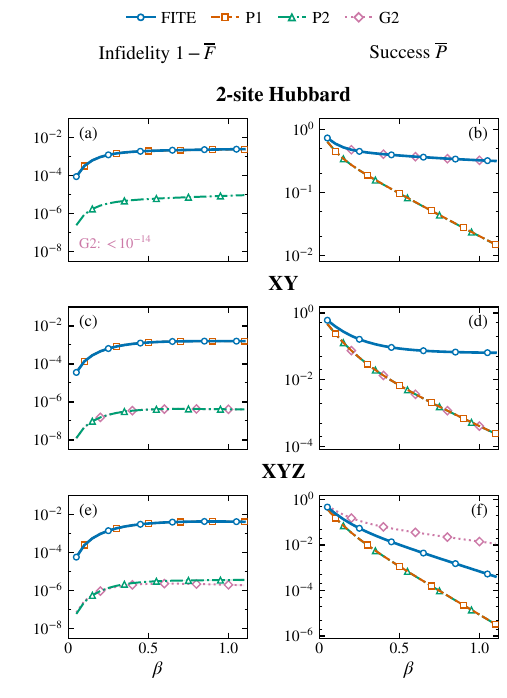}\caption{Panels (a,c,e) show mean infidelity with 95\% pointwise bands, and
panels (b,d,f) show exact Haar-mean success. The rows correspond to
Hubbard, XY and XYZ. All baselines use $\tau=0.05$, and FITE uses
phase-derived coefficients. Success axes have different ranges. Hubbard
G2 errors below $10^{-14}$ are omitted in (a), and overlapping curves
indicate equal results. Coefficient sampling noise is excluded.}\label{fig:accuracy-success}
\end{figure}

Figure \ref{fig:accuracy-success} compares mean infidelity and success
probability over $0.05\leq\beta\leq1.1$, using one fixed compiled
representation for FITE and a fixed step size $\tau=0.05$ for the
baselines. In panels (a,c,e), the FITE mean infidelity remains below
1\% throughout this interval. At $\beta=1.1$, it reaches 0.00245,
0.00158 and 0.00408 for Hubbard, XY and XYZ, respectively. These values
are close to those of P1, whereas P2 and G2 achieve lower infidelities
at this step size. For the two-qubit Hubbard model, G2 treats the
entire Hamiltonian as one block and is exact up to numerical precision.

Panels (b,d,f) show higher mean success for FITE than for P1 and P2
in all three models, with the relative gain increasing with $\beta$.
At $\beta=1.1$, FITE reaches mean success probabilities of 0.317,
0.0641 and $4.09\times10^{-4}$ for Hubbard, XY and XYZ, respectively.
These are approximately 21, 263 and 123 times the corresponding P1
values; the gains over P2 are similar. Thus, in these benchmarks,
FITE maintains accuracy close to P1 while substantially increasing
the probability of obtaining an accepted state.

The comparison with grouped filtering depends on the model. G2 has
nearly the same mean success as FITE for Hubbard and a higher mean
success for XYZ. For XY, the G2 and P2 curves overlap in both accuracy
and success, and FITE retains its success advantage. The improvement
over termwise PITE therefore does not extend uniformly to grouped
filters.

The absolute success probability for XYZ remains small even with FITE,
making the cost of repeated attempts relevant. Moreover, the methods
achieve different accuracies at fixed $\tau$. We therefore compare
gate costs at a common accuracy threshold, including failed attempts
before obtaining an accepted output.

\subsection{Resources at a common accuracy threshold}\label{sec:matched}

\begin{figure*}
\centering
\centering{}\includegraphics[width=1\textwidth]{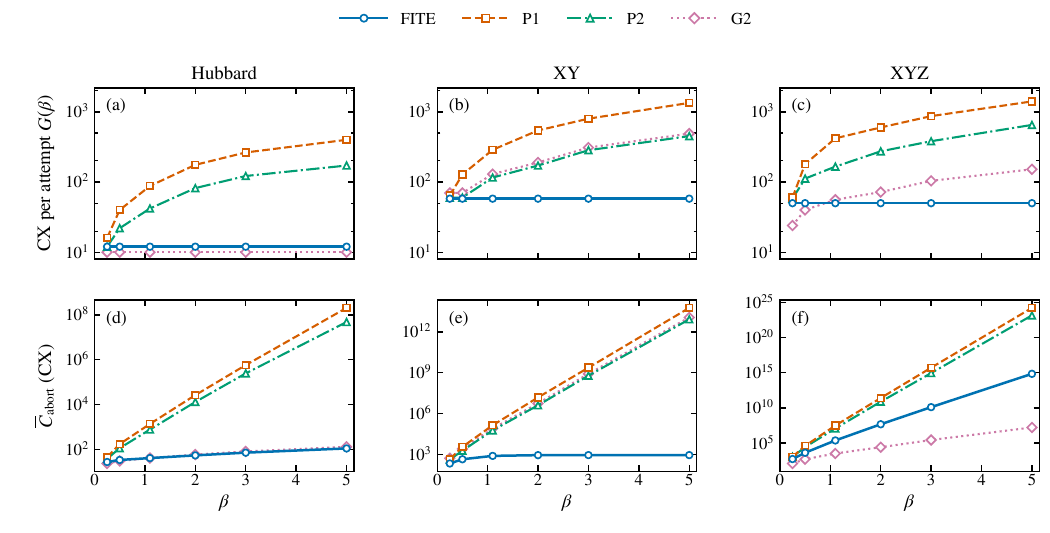}\caption{Online logical CX costs at a common 1\% mean-infidelity threshold.
Panels (a--c) show CX gates per full attempt, $G(\beta)$; panels
(d--f) show the input-averaged cost per accepted output, $\overline{C}_{{\rm abort}}$,
including early abort and retries of the same input. At each sampled
$\beta$, each baseline uses the smallest $M\in\{1,\ldots,400\}$
satisfying $(1-\overline{F})+1.96\,\mathrm{SE}\protect\leq0.01$;
FITE uses one fixed compiled representation. All circuits use one
reused ancilla. Counts exclude input preparation, offline compilation
and coefficient estimation, and hardware routing. The upper panels
share a vertical scale; the lower-panel ranges differ. Lines guide
the eye. See Appendix \ref{app:resources} for accounting details.}\label{fig:matched-resources}
\end{figure*}

Figure \ref{fig:matched-resources} compares online resource use at
a common accuracy requirement for six imaginary times between $\beta=0.25$
and 5. We impose an empirical mean-infidelity threshold $\epsilon=0.01$,
requiring $(1-\overline{F})+1.96\,\mathrm{SE}\leq\epsilon$, where
SE is the standard error of the estimated mean infidelity. At each
sampled $\beta$, we select the smallest passing baseline step count
from $M=1,\ldots,400$. FITE uses the same compiled representation
throughout and satisfies this criterion at all sampled times. The
criterion applies to the complex-Haar input ensemble, without guaranteeing
accuracy for every input.

We distinguish the gates needed for one complete circuit execution
from those needed to obtain an accepted output. All circuits use one
reused ancilla and the logical gate set $\{u,\mathrm{CX}\}$. The
count $G(\beta)$ gives the number of CX gates per full attempt. The
expected cost $C_{{\rm abort}}(\psi;\beta)$ includes repeated attempts
on the same input until success, stopping each failed attempt at the
first unsuccessful ancilla measurement. We report the input average
$\overline{C}_{{\rm abort}}=\mathbb{E}_{\psi}C_{{\rm abort}}(\psi;\beta)$.
Counts exclude input preparation and are taken before hardware routing
and additional circuit optimization.

Figure \ref{fig:matched-resources}(a--c) shows the cost per full
attempt. FITE uses 12, 58 and 50 CX for Hubbard, XY and XYZ, respectively,
independent of $\beta$: changing the evolution time changes only
the filter angles. The P1 and P2 counts increase as more steps are
needed to meet the accuracy threshold. For both chains, FITE uses
fewer CX than all three baselines at $\beta=1.1$ and 5. Hubbard G2
remains at 10 CX, since a single block implements exact ITE.

Including retries gives the costs in Fig. \ref{fig:matched-resources}(d--f).
FITE has a lower mean cost per accepted output than P1 and P2 at every
sampled time, and also outperforms G2 for XY. At $\beta=1.1$, Hubbard
FITE and G2 both cost approximately 41 CX per accepted output. For
XYZ, however, G2 costs $3.3\times10^{3}$ CX compared with $2.36\times10^{5}$
for FITE, despite using slightly more gates per full attempt (56 versus
50). The higher success probability of grouped filtering outweighs
this difference in circuit size.

The distinction becomes more pronounced at longer times: for XYZ,
the FITE cost per accepted output reaches $7.21\times10^{14}$ CX
at $\beta=5$, although its cost per full attempt remains 50 CX. A
fixed gate count therefore does not ensure a low cost per accepted
output. These online comparisons exclude offline compilation and coefficient
estimation, whose amortization depends on reuse. We next examine the
sources of the output error and its sensitivity to the compilation
settings.

\subsection{Accuracy versus compilation interval and ansatz depth}\label{sec:accuracy-checks}

For the main small-$\Delta t$ fits, phase conversion and physical-Hamiltonian
projection give similar output accuracy (Appendix \ref{app:coefficient-checks}).
Defining $H_{\mathrm{eff}}=(i/\Delta t)\log U_{\mathrm{T}}(\Delta t)$
using the principal matrix logarithm, we find that the target bias
dominates the fit residual.

\begin{figure*}
\centering
\centering{}\includegraphics[width=1\textwidth]{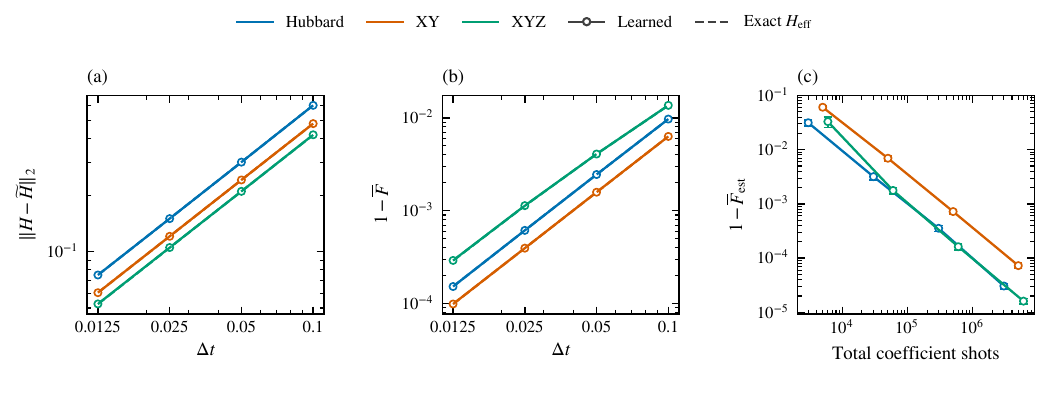}\caption{Panels (a,b) compare the dependence on the compilation interval for
learned phase coefficients (circles) and the exact $H_{\mathrm{eff}}$
(dashed lines); the dashed controls use $\widetilde{H}=H_{\mathrm{eff}}$.
Panel (c) shows $1-\overline{F}_{\mathrm{est}}$, the infidelity between
sampled and exact projection outputs, averaged over 512 sampling repetitions
and 128 Haar inputs. Error bars are $\pm1.96$ SE over repetitions.
Panels (b,c) use $\beta=1.1$. Appendix \ref{app:coefficient-estimation}
gives the sampling analysis and conservative shot bounds.}\label{fig:controls}
\end{figure*}

Figure \ref{fig:controls}(a,b) tests the dependence on the compilation
interval. We reduce $\Delta t$ and reoptimize the parameters while
keeping each model's ansatz depth and diagonal support fixed. At $\beta=1.1$,
reducing $\Delta t$ from 0.05 to 0.0125 lowers the phase-route mean
infidelities to $1.53\times10^{-4}$, $9.91\times10^{-5}$ and $2.92\times10^{-4}$
for Hubbard, XY and XYZ, respectively. Both the Hamiltonian error
and output infidelity closely follow the controls obtained from exact
$H_{{\rm eff}}$, supporting the attribution of the main errors to
the first-order Trotter target.

\begin{figure}
\centering
\centering{}\includegraphics[width=1\columnwidth]{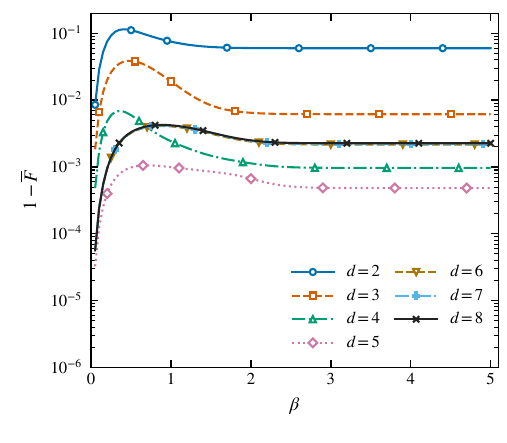}\caption{The XYZ depth scan uses $n=3$ and $\Delta t=0.05$. Curves show
mean infidelity; uncertainty bands are omitted for clarity. At $\beta=5$,
mean infidelity is $4.83\times10^{-4}$ for $d=5$ and 0.00226 for
$d=8$; the effective-target ground-state error is also 0.00226.}\label{fig:ansatz-errors}
\end{figure}

Increasing the basis depth at fixed $\Delta t$ has a different effect.
In the XYZ scan of Fig. \ref{fig:ansatz-errors}, the $d=8$ fit has
a lower compilation cost than the $d=5$ fit, but its mean infidelity
at $\beta=5$ is 0.00226, compared with $4.83\times10^{-4}$ for $d=5$.
The deeper fit more faithfully reproduces the biased target $U_{{\rm T}}$
and approaches the output error associated with $H_{{\rm eff}}$.
Consequently, reducing $C_{{\rm LHST}}$ alone does not guarantee
a more accurate approximation to physical ITE; the target error must
also be controlled.

\section{Discussion}\label{sec:discussion}

The numerical results distinguish improvements in success probability
from reductions in circuit cost per attempt. We now explain the mechanisms
behind these changes and discuss how accuracy, retries and offline
preparation determine their practical value.

\subsection{Origin of the success-probability gain}\label{sec:normalization}

For exact diagonalization $\widetilde{H}=H$, FITE gives the same
normalized output as termwise PITE in the $\tau\to0$ limit. Any difference
in success probability then comes from filter normalization. Maximal
local normalization of each Pauli filter gives

\begin{equation}
\ln\frac{P_{{\rm FITE}}(\psi)}{P_{{\rm PITE}}(\psi)}=2\beta(\lambda-\Lambda).\label{eq:ideal-gain}
\end{equation}

Thus $\Lambda<\lambda$ yields an exponential relative gain with $\beta$.
The ratio is input-independent, so the same relation applies to probabilities
averaged over the same ensemble. Appendix \ref{app:success} gives
the derivation and the corrections due to representation and product-formula
errors. Table \ref{tab:norm-reduction} shows that the fitted slopes
closely match the ideal predictions for all three benchmarks, supporting
this normalization mechanism.

\begin{table}
\centering
\centering{}\caption{Theoretical and fitted slopes of $\ln(\overline{P}_{{\rm FITE}}/\overline{P}_{{\rm P1}})$
versus $\beta$. Fits are constrained to pass through the origin over
$0.05\protect\leq\beta\protect\leq1.1$, with $\tau=0.05$ for P1.
The theoretical slopes $2(\lambda-\Lambda)$ use the main phase-derived
coefficients. The maximum absolute deviations from $2\beta(\lambda-\Lambda)$
are 0.0199, 0.0115 and 0.0235 for Hubbard, XY and XYZ, respectively.}
\label{tab:norm-reduction}{\small{}%
\begin{tabular}{lrrrr}
\hline 
Model & $\lambda$ & $\Lambda$ & Theoretical slope & Fitted slope\tabularnewline
\hline 
Hubbard & 5 & 3.6 & 2.7972 & 2.7789\tabularnewline
XY & 8 & 5.46 & 5.0757 & 5.0649\tabularnewline
XYZ & 10.5 & 8.3 & 4.3941 & 4.3708\tabularnewline
\hline 
\end{tabular}}
\end{table}

This comparison uses the termwise Pauli normalization of P1/P2. Grouping
terms can reduce the normalization loss of the baseline, explaining
the higher success of G2 for XYZ.

The open XY chain illustrates the size dependence of this mechanism:
its exact diagonal representation gives an ideal logarithmic gain
growing asymptotically as $\beta n$ (see Appendix \ref{app:xy-analysis}).
This scaling concerns relative success and does not establish efficient
compilation or large absolute success.

\subsection{Circuit cost per attempt}\label{sec:depth}

Commutativity allows the diagonal filters to cover the full imaginary
time without repeating product-formula steps. A complete FITE attempt
still includes two basis transformations and $K$ filters. Writing
$G$ for the logical CX count, the costs before additional circuit
optimization have the form

\begin{equation}
\begin{aligned}G_{\mathrm{FITE}} & =2G(W)+G(\mathrm{filters}),\\
G_{\mathrm{PITE}} & \sim MG(\mathrm{step}).
\end{aligned}
\label{eq:attempt-gates}
\end{equation}

Here the filter and step costs refer to the complete diagonal sequence
and one baseline step, respectively; the second relation allows for
merging adjacent filters. For fixed $W$ and retained diagonal support,
changing $\beta$ changes only filter angles. Both the gate count
and logical circuit depth per full attempt therefore remain fixed,
explaining the flat FITE curves in Fig. \ref{fig:matched-resources}(a--c).

A saving occurs when the basis transformations and diagonal filters
cost less than the baseline steps needed to meet the same accuracy
requirement. The comparison is restricted to evolution times for which
the fixed representation meets that requirement. The practical value
of this saving also depends on failed attempts and the offline cost
of obtaining the representation.

\subsection{Conditions for practical benefit}\label{sec:limits}

The combined benefit of the preceding improvements depends on the
total cost of obtaining accurate accepted outputs. This requires an
accurate representation with a sufficiently shallow basis circuit
$W$ and manageable diagonal support $K$. In the size tests, the
four- and five-qubit XYZ fits remain above 1\% mean infidelity under
the chosen ansatz and finite optimization budget (see Appendix \ref{app:size-dependence}
for details). These results therefore do not establish efficient scaling.
Accuracy must also be checked for the intended inputs: the XYZ example
in Appendix \ref{app:physical} shows how symmetry leakage can populate
a lower-energy sector absent from the input and distort the normalized
output.

For a usable representation, the online comparison is based on $\overline{C}_{{\rm abort}}$
at the required accuracy, including early termination of failed attempts
and retries of the same input. The XYZ comparison with G2 shows that
fewer gates per attempt can still lead to a higher cost per accepted
output. Moreover, small absolute success probabilities can make this
cost large even when the relative success gain is substantial.

Offline compilation and coefficient estimation must then be included
in the cost comparison. For a fixed Hamiltonian, the same representation
can be reused across input states and imaginary times within its validated
accuracy range. Such reuse can offset the initial investment if the
accumulated online savings are sufficiently large. The coefficient
precision must support the intended range of $\beta$. We have not
quantified the full offline cost, so the reuse needed for an overall
advantage remains undetermined. Finite-shot training, hardware noise
and routing also remain outside the present benchmarks.

\section{Conclusions}\label{sec:conclusions}

We have presented FITE, a probabilistic imaginary-time evolution scheme
based on a compiled approximate diagonal representation of the Hamiltonian.
The method determines a basis circuit offline and obtains the diagonal
coefficients directly from the physical Hamiltonian, avoiding energy-branch
ambiguities in the optimized real-time phases. The resulting commuting
filters cover the full imaginary time. For a fixed representation,
the circuit can therefore be reused across inputs and evolution times
with unchanged depth and gate count per attempt.

In the small-system benchmarks, FITE reduces the online CX cost per
accepted output relative to first- and second-order termwise Pauli
PITE at a common accuracy threshold, including failed attempts. Our
analysis relates the observed success gains to reduced filter-normalization
loss. The comparison with grouped filtering shows that the advantage
is model dependent and that a smaller circuit need not give a lower
cost per accepted output.

Practical benefit requires an accurate, compact representation and
sufficient reuse to offset compilation and coefficient-estimation
costs. The present results establish online savings for the tested
systems; efficient scaling and an overall cost advantage remain to
be demonstrated. Developing basis circuits that preserve Hamiltonian
symmetries and reducing the offline preparation cost are directions
for extending the method.
\begin{acknowledgments}
The authors used OpenAI's GPT model (gpt-6.1-sol) through Codex
to refine the wording of the manuscript.
\end{acknowledgments}

\appendix

\section{Circuit and optimization settings}\label{app:circuits}

\begin{figure}
\centering
\centering{}\includegraphics[width=1\columnwidth]{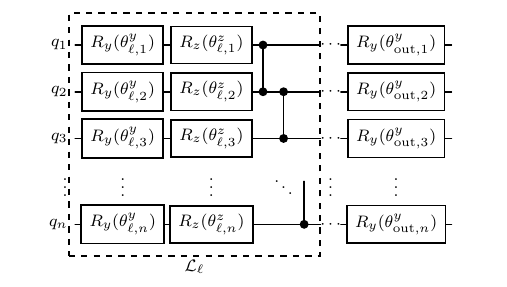}\caption{The $n$-qubit basis ansatz. The dashed box denotes layer $\mathcal{L}_{\ell}$,
with $\ell=1,\ldots,d$. Vertical and diagonal dots indicate omitted
qubits and CZ gates, while horizontal dots indicate the remaining
layers.}\label{fig:sm-default-ansatz}
\end{figure}

The numerical benchmarks use the common ansatz shown in Fig. \ref{fig:sm-default-ansatz}.
Each of its $d$ layers contains independent $R_{y}$ and $R_{z}$
rotations on every qubit, followed by CZ gates between nearest neighbors
in an open chain. A final layer of $R_{y}$ rotations completes the
circuit. This choice provides a common parametrization across the
models without imposing their individual symmetries.

For qubits $q=1,\ldots,n$, define $R_{a}^{(\ell)}=\bigotimes_{q=1}^{n}R_{a}(\theta_{\ell,q}^{a})$,
where $a\in\{y,z\}$, and $C_{Z}=\prod_{q=1}^{n-1}\mathrm{CZ}_{q,q+1}$.
The basis circuit is

\begin{equation}
W(\boldsymbol{\theta})=R_{y}^{(\mathrm{out})}\overleftarrow{\prod}_{\ell=1}^{d}\left[C_{Z}R_{z}^{(\ell)}R_{y}^{(\ell)}\right],\label{eq:sm-default-w}
\end{equation}

where $R_{y}^{(\mathrm{out})}=\bigotimes_{q=1}^{n}R_{y}(\theta_{\mathrm{out},q}^{y})$
and the arrow places the first layer on the right, so it acts first.
All rotation angles are independent, giving $N_{\theta}=n(2d+1)$
parameters and $d(n-1)$ CZ gates.

The diagonal unitary $D$ used in compilation and the subsequent filters
use the same retained strings. For the main configurations, these
are

\begin{equation}
\begin{aligned}\mathcal{S}_{Z}^{\mathrm{Hub}} & =\{Z_{1},Z_{2},Z_{1}Z_{2}\},\\
\mathcal{S}_{Z}^{\mathrm{XY}} & =\{Z_{q}\}_{q=1}^{n},\\
\mathcal{S}_{Z}^{\mathrm{XYZ}} & =\{Z_{q}\}_{q=1}^{n}\cup\{Z_{q}Z_{r}\}_{1\leq q<r\leq n}.
\end{aligned}
\label{eq:sm-default-d-support}
\end{equation}

These choices give $K=3$, $n$, and $n(n+1)/2$, respectively. The
Hubbard benchmark is restricted to one spin-up and one spin-down particle,
with one qubit per spin species labeling the occupied site. Hopping
$t=1$ and interaction $U=6$, after subtracting the constant $UI/2$,
give the two-qubit Hamiltonian in Eq. (\ref{eq:models}).

To implement the general diagonal filter in Fig. \ref{fig:fite-workflow}(c),
write $\mathcal{Z}_{k}=\prod_{j=1}^{w_{k}}Z_{q_{j}}$, where $w_{k}$
is the number of nonidentity factors. CX gates from $q_{1},\ldots,q_{w_{k}-1}$
encode their parity onto $q_{w_{k}}$. For nonzero $\xi_{k}$, applying
$X^{b_{k}}$ to this qubit, with $b_{k}=[1+\operatorname{sgn}(\xi_{k})]/2$,
selects the eigenspace to be attenuated. This system qubit controls
an $R_{y}$ rotation on an ancilla initialized in $\ket 0$, with
angle

\begin{equation}
\gamma_{k}=2\arccos\!\left(e^{-2\beta|\xi_{k}|}\right).\label{eq:sm-filter-angle}
\end{equation}

For nonzero $\xi_{k}$, the control activates the rotation in the
$\mathcal{Z}_{k}$ eigenspace with eigenvalue $\operatorname{sgn}(\xi_{k})$;
a zero coefficient gives the identity filter. Postselecting the ancilla
on zero attenuates that eigenspace by $\cos(\gamma_{k}/2)=e^{-2\beta|\xi_{k}|}$
and leaves the opposite eigenspace unchanged. Applying $X^{b_{k}}$
again and uncomputing the parity therefore yields $\mathcal{K}_{k}$
in Eq. (\ref{eq:diagonal-kraus}). The filter angle $\gamma_{k}$
depends on the requested $\beta$ and is distinct from the real-time
training angle $\varphi_{k}$. A complete successful attempt uses
$K$ ancilla measurements and $K-1$ resets, reusing one ancilla throughout.

The compilation procedure applies to all three model families. For
a specified Hamiltonian, basis ansatz and diagonal support, we minimize
$C_{\mathrm{LHST}}$ jointly over $\boldsymbol{\theta}$ and $\boldsymbol{\varphi}$
using L-BFGS-B, a limited-memory quasi-Newton algorithm for bound-constrained
optimization \citep{Byrd1995}. Independent searches initialize each
angle from $0.1\mathcal{N}(0,1)$ and constrain the angles to $[-\pi,\pi]$.
The cost is evaluated without measurement noise, and we select the
retained fit with the lowest compilation cost for each configuration.

The main Hubbard and XY benchmarks use previously optimized solutions.
For the main XYZ configuration $(n,d,K)=(3,8,6)$, additional independent
searches allow up to 1500 initializations, with at most 1000 iterations
per initialization, and stop after a batch reaches $C_{\mathrm{LHST}}<10^{-7}$.
This search uses 1152 initializations. Selected candidates are further
optimized for up to 10,000 iterations under the same angle bounds.

For the compilation-interval scan in Fig. \ref{fig:controls}(a,b),
each model starts from a previously obtained fit at $\Delta t_{0}=0.05$.
We retain its basis angles and rescale the initial diagonal angles
by $\Delta t/\Delta t_{0}$, then reoptimize both sets of parameters
with L-BFGS-B. These continuation runs allow up to 1800 iterations
without angle bounds.

\section{Coefficient estimation and sampling cost}\label{app:coefficient-estimation}

We quantify the sampling cost of physical-Hamiltonian coefficient
recovery, keeping $W$ and the retained diagonal support fixed. Throughout
this appendix, $\xi_{k}=\xi_{k}^{\mathrm{proj}}$ denotes the exact
projection coefficient and $\lambda=\sum_{j}|c_{j}|$. For each shot,
sample $P_{j}$ with probability $|c_{j}|/\lambda$ and run the trace
test in Fig. \ref{fig:coefficient-trace-test}. If the ancilla outcome
is $b\in\{0,1\}$, assign the value $V_{k}=\lambda\operatorname{sgn}(c_{j})(-1)^{b}$.
For ideal gates and measurements, Eq. (\ref{eq:coefficient-readout})
gives $\mathbb{E}[V_{k}]=\xi_{k}$. Since $V_{k}^{2}=\lambda^{2}$,
the sample mean $\widehat{\xi}_{k}=S_{k}^{-1}\sum_{s=1}^{S_{k}}V_{k,s}$
from $S_{k}$ independent shots has variance $(\lambda^{2}-\xi_{k}^{2})/S_{k}$.

Hoeffding's inequality \citep{Hoeffding1963} bounds the probability
of a coefficient error exceeding $\eta/K$ by $2\exp[-S_{k}\eta^{2}/(2\lambda^{2}K^{2})]$.
A union bound over the $K$ coefficients therefore gives the sufficient
requirement

\begin{equation}
S_{k}\geq\frac{2\lambda^{2}K^{2}}{\eta^{2}}\log\frac{2K}{\alpha},\qquad k=1,\ldots,K,\label{eq:coefficient-shots}
\end{equation}

which ensures $\Pr(\|\widehat{\boldsymbol{\xi}}-\boldsymbol{\xi}\|_{1}>\eta)\leq\alpha$
for tolerance $\eta>0$ and failure probability $0<\alpha<1$.

This coefficient bound also controls the error in the normalized output.
On the event $\|\widehat{\boldsymbol{\xi}}-\boldsymbol{\xi}\|_{1}\leq\eta$,
the perturbation $\delta\mathcal{H}=\sum_{k}(\widehat{\xi}_{k}-\xi_{k})\mathcal{Z}_{k}$
satisfies $\|\delta\mathcal{H}\|_{2}\leq\eta$. Because the diagonal
terms commute, the sampled output is obtained from the exact projection
output by applying $T=e^{-\beta\delta\mathcal{H}}$ in the diagonal
basis and renormalizing. For $\beta\geq0$, the eigenvalues of $T$
lie between $a_{-}=e^{-\beta\eta}$ and $a_{+}=e^{\beta\eta}$. Let
$\ket{\chi}$ denote the normalized exact projection output in that
basis. Using $T^{2}\leq(a_{-}+a_{+})T-a_{-}a_{+}I$ gives

\begin{equation}
F_{\mathrm{est}}=\frac{\langle T\rangle_{\chi}^{2}}{\langle T^{2}\rangle_{\chi}}\geq\frac{4a_{-}a_{+}}{(a_{-}+a_{+})^{2}}=\frac{1}{\cosh^{2}(\beta\eta)}.\label{eq:coefficient-fidelity-bound}
\end{equation}

Hence $1-F_{\mathrm{est}}\leq\tanh^{2}(\beta\eta)$ for every input
whenever the coefficient bound holds. This compares outputs formed
with sampled and exact projection coefficients; errors from the compiled
basis or omitted diagonal terms remain outside this estimate.

The trace test requires $2n+1$ qubits. With the default layered ansatz,
its logical CX count per shot is

\begin{equation}
G_{kj}^{\mathrm{trace}}=2d(n-1)+n+\mathrm{wt}(P_{j})+\mathrm{wt}(\mathcal{Z}_{k}),\label{eq:coefficient-shot-gates}
\end{equation}

where $\mathrm{wt}$ counts nonidentity sites. The first term comes
from the unconditional $W$ and $W^{\dagger}$ circuits, the second
from Bell-pair preparation, and the remaining terms from the controlled
Pauli strings. This count uses the $\{u,\mathrm{CX}\}$ gate set without
routing or cancellations between circuit blocks.

Table \ref{tab:estimator-cost} evaluates the sufficient total $N_{\mathrm{shots}}=\sum_{k}S_{k}$
for the main configurations. Taking $\alpha=0.05$ and $\eta=\operatorname{arctanh}(\sqrt{10^{-3}})/5$
bounds the sampling contribution to infidelity by $10^{-3}$ for all
inputs and $0\leq\beta\leq5$, with 95\% confidence. These are conservative
sufficient budgets. Figure \ref{fig:controls}(c) instead reports
mean sampling errors at $\beta=1.1$, using $\overline{F}_{\mathrm{est}}=\mathbb{E}_{\mathrm{shots}}\mathbb{E}_{\psi}[F_{\mathrm{est}}]$,
estimated from 512 sampling repetitions and 128 Haar inputs. Coefficient
readout is an offline cost, excluded from the online resource counts.

\begin{table}
\centering
\centering{}\caption{Coefficient-estimation resources for the main configurations. The
total shot bound in Eq. (\ref{eq:coefficient-shots}) guarantees sampling
infidelity at most $10^{-3}$ for all inputs and $0\protect\leq\beta\protect\leq5$
with 95\% confidence. CX counts are averaged over sampled Hamiltonian
terms; each range spans the retained diagonal strings.}
\label{tab:estimator-cost}{\small{}%
\begin{tabular}{lrrr}
\hline 
Model & Qubits & Mean CX/shot & $N_{{\rm shots}}^{{\rm bound}}$\tabularnewline
\hline 
Hubbard & 5 & 8.6--9.6 & $1.61\times10^{8}$\tabularnewline
XY & 11 & 56 & $2.12\times10^{9}$\tabularnewline
XYZ & 7 & 37.6--38.6 & $6.52\times10^{9}$\tabularnewline
\hline 
\end{tabular}}
\end{table}

\section{Success-probability analysis}\label{app:success}

We derive the ideal success-probability gain by collecting the local
filter normalization factors. For P1, each factor $e^{-\tau c_{j}P_{j}}$
is multiplied by $e^{-\tau|c_{j}|}$ to have unit operator norm. With
$\tau=\beta/M$, the normalization over all $M$ steps is $e^{-\beta\lambda}$,
where $\lambda=\sum_{j}|c_{j}|$. Thus,

\begin{equation}
\begin{aligned}\mathcal{K}_{\mathrm{PITE}}(\beta,\tau) & =e^{-\beta\lambda}\mathcal{T}_{\tau}(\beta),\\
\mathcal{T}_{\tau}(\beta) & =\left[\overleftarrow{\prod}_{j=1}^{N_{P}}e^{-\tau c_{j}P_{j}}\right]^{M}.
\end{aligned}
\label{eq:sm-pite-kraus}
\end{equation}

The arrow fixes the circuit order. Symmetric P2 has the same scalar
normalization because its half steps give each Pauli term the same
total evolution time $\beta$. The following formulas also apply to
P2 with $\mathcal{T}_{\tau}$ replaced by its symmetric product formula.
Varying $M$ changes the product-formula error but leaves the scalar
attenuation unchanged at fixed $\beta$. For a normalized input $\ket{\psi}$,
the success probability is

\begin{equation}
P_{\mathrm{PITE}}(\psi)=e^{-2\beta\lambda}\bra{\psi}\mathcal{T}_{\tau}^{\dagger}(\beta)\mathcal{T}_{\tau}(\beta)\ket{\psi}.\label{eq:sm-pite-probability}
\end{equation}

Dividing the FITE probability in Eq. (\ref{eq:fite-success-probability})
by this expression cancels the input-dependent norm when $\widetilde{H}=H$
and $\tau\to0$ at fixed $\beta$. This gives Eq. (\ref{eq:ideal-gain});
its input independence also makes it valid for probabilities averaged
over any common input ensemble.

For an approximate representation or finite $\tau$, the same comparison
gives the exact decomposition

\begin{equation}
\begin{aligned}\ln\frac{P_{\mathrm{FITE}}(\psi)}{P_{\mathrm{PITE}}(\psi)} & =2\beta(\lambda-\Lambda)+\ln R_{\psi},\\
R_{\psi} & =\frac{\bra{\psi}e^{-2\beta\widetilde{H}}\ket{\psi}}{\bra{\psi}\mathcal{T}_{\tau}^{\dagger}\mathcal{T}_{\tau}\ket{\psi}}.
\end{aligned}
\label{eq:sm-state-correction}
\end{equation}

Here $\mathcal{T}_{\tau}=\mathcal{T}_{\tau}(\beta)$. The correction
$\ln R_{\psi}$ contains both representation and product-formula errors
and generally depends on the input. It vanishes in the ideal limit;
at finite errors, a gain requires the full right-hand side to be positive.

For the complex-Haar ensemble, the average input density operator
is $I/2^{n}$. Taking traces therefore gives the mean-probability
relation

\begin{equation}
\ln\frac{\overline{P}_{\mathrm{FITE}}}{\overline{P}_{\mathrm{PITE}}}=2\beta(\lambda-\Lambda)+\ln\!\left[\frac{\operatorname{Tr}(e^{-2\beta\widetilde{H}})}{\operatorname{Tr}(\mathcal{T}_{\tau}^{\dagger}\mathcal{T}_{\tau})}\right].\label{eq:sm-haar-enhancement}
\end{equation}

At finite errors, this ratio of mean probabilities need not equal
the average of the individual probability ratios. Evaluating the traces
gives the data for Table \ref{tab:norm-reduction}, including the
correction to the ideal linear dependence on $\beta$, without sampling
input states. These relations use termwise Pauli normalization for
P1/P2; grouped filters such as G2 have different normalization factors.

\section{Analytical results for the open XY chain}\label{app:xy-analysis}

For the open XY chain in Eq. (\ref{eq:models}), take a general coupling
$J_{\mathrm{XY}}\ne0$; the benchmark uses $J_{\mathrm{XY}}=-1$.
The original Pauli coefficient norm is $\lambda_{n}=2(n-1)|J_{\mathrm{XY}}|$.
A Jordan--Wigner transformation followed by a sine transform gives
independent fermionic modes $a_{q}$ with energies $4J_{\mathrm{XY}}\cos k_{q}$,
where $k_{q}=q\pi/(n+1)$. Using $a_{q}^{\dagger}a_{q}=(I-Z_{q}^{(a)})/2$
and $\sum_{q}\cos k_{q}=0$ yields the exact diagonal representation

\begin{equation}
\begin{aligned}H_{\mathrm{XY}} & =W\mathcal{H}_{\mathrm{XY}}W^{\dagger},\\
\mathcal{H}_{\mathrm{XY}} & =-2J_{\mathrm{XY}}\sum_{q=1}^{n}\cos k_{q}\,Z_{q}^{(a)}.
\end{aligned}
\label{eq:sm-xy-diagonal-hamiltonian}
\end{equation}

Its coefficient norm is

\begin{equation}
\Lambda_{n}^{\mathrm{opt}}=2|J_{\mathrm{XY}}|\sum_{q=1}^{n}|\cos k_{q}|.\label{eq:sm-xy-lambda-diagonal}
\end{equation}

The independent mode occupations realize extremal energies $\pm\Lambda_{n}^{\mathrm{opt}}$.
This representation therefore saturates the bound $\Lambda\geq\|H\|_{2}$
for exact diagonal representations, establishing the optimality of
its coefficient norm.

Evaluating the cosine sum with $x_{n}=\pi/[2(n+1)]$ gives the finite-size
reduction

\begin{equation}
\lambda_{n}-\Lambda_{n}^{\mathrm{opt}}=2|J_{\mathrm{XY}}|\begin{cases}
n-\csc x_{n}, & n\ \mathrm{even},\\
n-\cot x_{n}, & n\ \mathrm{odd}.
\end{cases}\label{eq:sm-xy-parity-reduction}
\end{equation}

For large $n$, expanding this finite-size reduction and applying
the ideal gain relation, Eq. (\ref{eq:ideal-gain}), gives

\begin{equation}
\begin{aligned}\lambda_{n}-\Lambda_{n}^{\mathrm{opt}} & =\left(2-\frac{4}{\pi}\right)|J_{\mathrm{XY}}|n-\frac{4}{\pi}|J_{\mathrm{XY}}|\\
 & \quad+O\!\left(\frac{|J_{\mathrm{XY}}|}{n}\right),\\
\ln\frac{P_{\mathrm{FITE}}}{P_{\mathrm{PITE}}} & \sim2\beta\left(2-\frac{4}{\pi}\right)|J_{\mathrm{XY}}|n.
\end{aligned}
\label{eq:sm-xy-asymptotic-success}
\end{equation}

The probability ratio assumes exact diagonalization and the vanishing-step
limit of termwise PITE.

For the five-site benchmark with $|J_{\mathrm{XY}}|=1$,

\begin{equation}
\begin{aligned}\lambda_{5} & =8,\qquad\Lambda_{5}^{\mathrm{opt}}=2(1+\sqrt{3})\simeq5.46410,\\
\lambda_{5}-\Lambda_{5}^{\mathrm{opt}} & =6-2\sqrt{3}\simeq2.53590.
\end{aligned}
\label{eq:sm-xy-n5}
\end{equation}

The phase-derived fit in Table \ref{tab:norm-reduction} gives $\lambda-\Lambda=2.53786$;
its deviation from the exact value includes Trotter error in the training
target and variational error.

Since $E_{0}=-\Lambda_{n}^{\mathrm{opt}}$, even optimal normalization
gives $\overline{P}\to g/2^{n}$ for Haar inputs as $\beta\to\infty$,
where the ground-state degeneracy is $g=1$ for even $n$ and 2 for
odd $n$. The relative gain therefore does not remove the limitation
from small absolute success.

\section{Validation of diagonal-coefficient recovery}\label{app:coefficient-checks}

We check coefficient recovery for the main Hubbard, XY and XYZ configurations
while keeping the compiled basis $W$ and retained diagonal support
fixed. Table \ref{tab:coefficients} compares the phase-derived coefficients
$\xi_{k}^{\mathrm{phase}}$ with the physical-Hamiltonian projection
coefficients $\xi_{k}^{\mathrm{proj}}$.

\begin{table}
\centering
\centering{}\caption{The table compares diagonal coefficients from phase conversion and
physical-Hamiltonian trace projection for the main configurations.
Both routes use the same compiled $W$ and retained Pauli support.}
\label{tab:coefficients}{\small{}%
\begin{tabular}{lrrr}
\hline 
Model & $\mathcal{Z}_{k}$ & $\xi_{k}^{\mathrm{phase}}$ & $\xi_{k}^{\mathrm{proj}}$\tabularnewline
\hline 
Hubbard & $Z_{1}$ & 0.30069 & 0.296507\tabularnewline
Hubbard & $Z_{2}$ & $-5.38730\times10^{-7}$ & $-5.83977\times10^{-14}$\tabularnewline
Hubbard & $Z_{1}Z_{2}$ & 3.30069 & 3.29651\tabularnewline
XY & $Z_{1}$ & -1.73169 & -1.72808\tabularnewline
XY & $Z_{2}$ & 0.999377 & 1.00312\tabularnewline
XY & $Z_{3}$ & $-8.54000\times10^{-9}$ & $-2.74140\times10^{-12}$\tabularnewline
XY & $Z_{4}$ & -0.99938 & -1.00312\tabularnewline
XY & $Z_{5}$ & -1.73169 & -1.72809\tabularnewline
XYZ & $Z_{1}$ & 1.30048 & 1.30048\tabularnewline
XYZ & $Z_{2}$ & 1.30132 & 1.3013\tabularnewline
XYZ & $Z_{1}Z_{2}$ & 2.20265 & 2.20263\tabularnewline
XYZ & $Z_{3}$ & 1.29916 & 1.29917\tabularnewline
XYZ & $Z_{1}Z_{3}$ & -1.00001 & -0.996683\tabularnewline
XYZ & $Z_{2}Z_{3}$ & 1.19935 & 1.19602\tabularnewline
\hline 
\end{tabular}}
\end{table}

At $\beta=1.1$, replacing $\xi_{k}^{\mathrm{phase}}$ by $\xi_{k}^{\mathrm{proj}}$
gives mean infidelities of 0.00247, 0.00158 and 0.00408 for Hubbard,
XY and XYZ, respectively, close to the phase-based values. Projection
recovers the retained diagonal components of $W^{\dagger}HW$, while
errors from off-diagonal components and omitted diagonal terms remain.

For the phase-derived representation, the small fit residuals $\|H_{\mathrm{eff}}-\widetilde{H}\|_{2}$
in Table \ref{tab:error-budget} support the selected phase branch.
The target bias $\|H-H_{\mathrm{eff}}\|_{2}$ dominates all three
main fits at $\Delta t=0.05$.

\begin{table}
\centering
\centering{}\caption{For the phase route at $\Delta t=0.05$, the table separates target
error $\delta_{\mathrm{T}}=\|H-H_{{\rm eff}}\|_{2}$, fit error $\delta_{{\rm fit}}=\|H_{{\rm eff}}-\widetilde{H}\|_{2}$
and total error $\delta=\|H-\widetilde{H}\|_{2}$. Projection-route
infidelity is evaluated at $\beta=1.1$. The maximum omitted diagonal
norm is $2.77\times10^{-6}$.}
\label{tab:error-budget}{\small{}%
\begin{tabular}{lrrrr}
\hline 
Model & $\delta_{\mathrm{T}}$ & $\delta_{{\rm fit}}$ & $\delta$ & $1-\overline{F}_{{\rm proj}}$\tabularnewline
\hline 
Hubbard & 0.301 & $2.96\times10^{-6}$ & 0.301 & 0.00247\tabularnewline
XY & 0.241 & 0.00373 & 0.241 & 0.00158\tabularnewline
XYZ & 0.21 & $1.46\times10^{-5}$ & 0.21 & 0.00408\tabularnewline
\hline 
\end{tabular}}
\end{table}

For every retained $\mathcal{Z}_{k}$ and Hamiltonian term $P_{j}$,
we evaluate the ancilla probability difference $p_{0}^{(kj)}-p_{1}^{(kj)}$
in Fig. \ref{fig:coefficient-trace-test} without sampling and compare
it with the corresponding normalized matrix trace. The maximum absolute
discrepancy across the three models is below $1.5\times10^{-15}$,
confirming agreement between the circuit readout and direct matrix
evaluation.

\begin{table*}
\centering
\centering{}\caption{Resource details for Fig. \ref{fig:matched-resources} at the common
1\% mean-infidelity threshold. All quantities refer to $\beta=1.1$
except $G(5)$, which uses the baseline step counts selected at $\beta=5$.
Here, $M$ is the step count, $G$ the CX count per full attempt,
$D_{2}$ the two-qubit depth, $\overline{P}$ the mean success probability,
and $\mathbb{E}_{\psi}[P^{-1}]$ the mean number of attempts for fixed
inputs. The mean cost $\overline{C}_{{\rm abort}}$ includes early
abort and same-input retries. Input preparation, measurement and offline
preparation costs are excluded.}
\label{tab:matched}{\small{}%
\begin{tabular}{lrrrrrrrr}
\hline 
Model & Method & $M$ & $G(1.1)$ & $G(5)$ & $D_{2}$ & $\overline{P}$ & $\mathbb{E}_{\psi}[P^{-1}]$ & $\overline{C}_{{\rm abort}}$\tabularnewline
\hline 
Hubbard & FITE & 1 & 12 & 12 & 12 & 0.317 & 5.41 & 41.1\tabularnewline
Hubbard & P1 & 11 & 88 & 400 & 88 & 0.0155 & 112 & $1.37\times10^{3}$\tabularnewline
Hubbard & P2 & 4 & 42 & 172 & 42 & 0.0181 & 99.7 & 765\tabularnewline
Hubbard & G2 & 1 & 10 & 10 & 10 & 0.316 & 5.43 & 41\tabularnewline
XY & FITE & 1 & 58 & 58 & 35 & 0.0641 & 27 & 784\tabularnewline
XY & P1 & 9 & 288 & 1344 & 272 & $2.53\times10^{-4}$ & $6.84\times10^{3}$ & $1.38\times10^{5}$\tabularnewline
XY & P2 & 2 & 116 & 452 & 116 & $3.08\times10^{-4}$ & $5.48\times10^{3}$ & $5.87\times10^{4}$\tabularnewline
XY & G2 & 2 & 130 & 490 & 130 & $3.08\times10^{-4}$ & $5.48\times10^{3}$ & $7.92\times10^{4}$\tabularnewline
XYZ & FITE & 1 & 50 & 50 & 49 & $4.09\times10^{-4}$ & $1.20\times10^{4}$ & $2.36\times10^{5}$\tabularnewline
XYZ & P1 & 14 & 420 & 1410 & 393 & $3.41\times10^{-6}$ & $1.44\times10^{6}$ & $3.17\times10^{7}$\tabularnewline
XYZ & P2 & 3 & 166 & 652 & 160 & $4.77\times10^{-6}$ & $1.03\times10^{6}$ & $1.20\times10^{7}$\tabularnewline
XYZ & G2 & 3 & 56 & 152 & 56 & 0.0131 & 341 & $3.3\times10^{3}$\tabularnewline
\hline 
\end{tabular}}
\end{table*}

\section{Baselines and resource accounting}\label{app:resources}

The Pauli baseline P1 applies first-order filters in the same term
order as the real-time target. P2 uses half steps in forward order,
a full step for the last term, and half steps in reverse order. Adjacent
filters for the same Pauli operator are merged \citep{Xie2024,Leadbeater2024}.

G2 uses the same symmetric schedule for nearest-neighbor two-site
blocks. Onsite fields are shared equally among the incident bonds,
and each block filter is normalized using its lowest eigenvalue \citep{Xie2024}.
For Hubbard, the entire Hamiltonian forms one block, so G2 implements
exact ITE.

The step count $M$ refers to imaginary-time steps regardless of formula
order; FITE is listed with $M=1$ for its combined filter sequence.
Figure \ref{fig:accuracy-success} fixes $\tau=0.05$ for the baselines.
For the resource comparison in Fig. \ref{fig:matched-resources},
we instead select the smallest $M$ passing the empirical 1\% mean-infidelity
criterion defined in Sec. \ref{sec:matched}.

At $\beta=1.1$, Table \ref{tab:matched} lists the selected step
counts, CX counts, two-qubit depths, success probabilities and retry
costs. It also includes the CX counts at $\beta=5$, using the step
counts selected at that time. Table \ref{tab:additional-counts} adds
single-qubit gates, quantum depth, measurements and resets for the
same configurations at $\beta=1.1$.

All counts use one reused ancilla and the logical gate set $\{u,\mathrm{CX}\}$.
They are taken before hardware routing and additional circuit optimization,
and exclude input preparation. The depths $D_{q}$ and $D_{2}$ are
measured in layers of all quantum gates and of two-qubit gates, respectively;
measurement and reset durations are excluded.

\begin{table}
\centering
\centering{}\caption{The table supplements Table \ref{tab:matched} at $\beta=1.1$ with
single-qubit gates $N_{1q}$, quantum depth $D_{q}$, measurements
$N_{m}$ and resets $N_{r}$ for the same configurations. All circuits
use one auxiliary qubit, and $D_{q}$ excludes measurement and reset
durations.}
\label{tab:additional-counts}{\small{}%
\begin{tabular}{lrrrrr}
\hline 
Model & Method & $N_{1q}$ & $D_{q}$ & $N_{m}$ & $N_{r}$\tabularnewline
\hline 
Hubbard & FITE & 38 & 36 & 3 & 2\tabularnewline
Hubbard & P1 & 132 & 176 & 33 & 32\tabularnewline
Hubbard & P2 & 68 & 85 & 17 & 16\tabularnewline
Hubbard & G2 & 20 & 21 & 1 & 0\tabularnewline
XY & FITE & 238 & 96 & 5 & 4\tabularnewline
XY & P1 & 432 & 480 & 72 & 71\tabularnewline
XY & P2 & 174 & 203 & 29 & 28\tabularnewline
XY & G2 & 260 & 273 & 13 & 12\tabularnewline
XYZ & FITE & 188 & 130 & 6 & 5\tabularnewline
XYZ & P1 & 728 & 815 & 126 & 125\tabularnewline
XYZ & P2 & 284 & 333 & 49 & 48\tabularnewline
XYZ & G2 & 112 & 119 & 7 & 6\tabularnewline
\hline 
\end{tabular}}
\end{table}

At fixed $\beta$, repeating full attempts on the same input costs
$C_{{\rm full}}(\psi)=G/P(\psi)$ per accepted output. When an attempt
stops at the first failed filter, its expected CX cost $A(\psi)$
is obtained by weighting each circuit segment by the probability of
reaching it. The corresponding cost is $C_{{\rm abort}}(\psi)=A(\psi)/P(\psi)$,
and we report $\overline{C}_{{\rm abort}}=\mathbb{E}_{\psi}[A(\psi)/P(\psi)]$.
Drawing a new input after every failure would instead give $G/\overline{P}$
for full attempts and change the accepted ensemble.

For complex-Haar input states, the squared amplitudes in an eigenbasis
of $\mathcal{K}^{\dagger}\mathcal{K}$ can be generated by normalizing
independent exponential variables. Writing $\mu_{a}$ for the eigenvalues
gives

\begin{equation}
\mathbb{E}_{\psi}[P(\psi)^{-1}]=(2^{n}-1)\int_{0}^{\infty}\prod_{a=1}^{2^{n}}(1+t\mu_{a})^{-1}\,dt,\label{eq:inverse-success}
\end{equation}

where $t$ is an auxiliary integration variable. The mean cost of
full attempts is therefore $\overline{C}_{{\rm full}}=G\,\mathbb{E}_{\psi}[P(\psi)^{-1}]$.
We obtain $\overline{C}_{{\rm abort}}$ from the analogous integral
for the ratio $A(\psi)/P(\psi)$.

\section{System-size dependence}\label{app:size-dependence}

Table \ref{tab:sizes} examines the dependence on system size using
XY chains with $n=2,\ldots,5$ and XYZ chains with $n=3,4,5$. All
configurations use $\Delta t=0.05$ and phase-derived coefficients.
We set $\beta\Delta=2$, where $\Delta$ is the energy gap above the
full ground manifold, to compare evolution at a common time relative
to the inverse gap. For XYZ, the tested configurations vary the basis
depth and compare support restricted to single- and two-qubit $Z$
strings with the full diagonal support $K=2^{n}-1$. We average over
the same complex-Haar input ensembles as in the main benchmarks and
report logical CX counts per full attempt.

For the four- and five-qubit XYZ configurations, independent searches
allow up to 1500 initializations per configuration, with at most 1000
iterations per initialization, and stop after a batch reaches $C_{\mathrm{LHST}}<10^{-7}$.
Each configuration uses the full budget of 1500 initializations. Selected
candidates are further optimized for up to 10,000 iterations with
the angles constrained to $[-\pi,\pi]$.

\begin{table*}
\centering
\centering{}\caption{The size tests use $\Delta t=0.05$, $\beta\Delta=2$, and phase coefficients
from the retained fit with the lowest compilation cost for each configuration.
Infidelity and success are averaged over complex-Haar inputs, and
$G$ is the logical CX count per full attempt.}
\label{tab:sizes}{\small{}%
\begin{tabular}{lrrrrrrr}
\hline 
Model & $n$ & $d$ & $K$ & $C_{{\rm LHST}}$ & $1-\overline{F}$ & $\overline{P}$ & $G$\tabularnewline
\hline 
XY & 2 & 2 & 2 & 0 & $2.39\times10^{-13}$ & 0.259 & 8\tabularnewline
XY & 3 & 3 & 3 & 0 & $8.83\times10^{-4}$ & 0.259 & 18\tabularnewline
XY & 4 & 6 & 4 & $4.00\times10^{-12}$ & $4.58\times10^{-4}$ & 0.0648 & 44\tabularnewline
XY & 5 & 6 & 5 & $4.08\times10^{-9}$ & 0.00159 & 0.0649 & 58\tabularnewline
XYZ & 3 & 8 & 6 & $1.20\times10^{-13}$ & 0.00418 & 0.00232 & 50\tabularnewline
XYZ & 4 & 12 & 15 & $8.01\times10^{-4}$ & 0.94 & $3.23\times10^{-25}$ & 136\tabularnewline
XYZ & 4 & 7 & 10 & 0.00213 & 0.939 & $7.78\times10^{-22}$ & 74\tabularnewline
XYZ & 4 & 7 & 15 & 0.00213 & 0.939 & $7.24\times10^{-22}$ & 106\tabularnewline
XYZ & 5 & 16 & 31 & 0.00128 & 0.0194 & $2.07\times10^{-6}$ & 288\tabularnewline
XYZ & 5 & 7 & 15 & 0.00273 & 0.0674 & 0.0043 & 106\tabularnewline
XYZ & 5 & 7 & 31 & 0.00243 & 0.0566 & $4.86\times10^{-6}$ & 216\tabularnewline
\hline 
\end{tabular}}
\end{table*}

The large infidelities of the four-qubit XYZ fits reflect errors in
the low-energy states. Here the gap is $\Delta\simeq0.3846$, so the
condition $\beta\Delta=2$ gives $\beta\simeq5.20$, compared with
approximately 0.77 and 0.81 for the three- and five-qubit XYZ chains.
For $(n,d,K)=(4,12,15),(4,7,10),(4,7,15)$, respectively, the ground
state of $\widetilde{H}$ has squared overlap $4.0\times10^{-10}$,
0.00399, and 0.00402 with the ground state of $H$, but squared overlap
0.9954, 0.9887, and 0.9887 with its first excited state. Thus, the
learned evolution preferentially amplifies a state close to the wrong
eigenstate of $H$, and the longer imaginary time makes this mismatch
more pronounced.

The small numerical values of $C_{\mathrm{LHST}}$ do not imply accurate
imaginary-time evolution: this cost measures the unitary fit at the
short training interval $\Delta t=0.05$. For the same three configurations,
$\|H-\widetilde{H}\|_{2}$ is approximately 1.46, 2.40, and 2.40,
all exceeding the physical gap. As a control, normalized evolution
under the exact effective target $H_{\mathrm{eff}}$ gives mean infidelity
$1.27\times10^{-3}$ at the same $\beta$ and on the same input ensemble.
The much larger errors of the retained fits therefore arise primarily
from the learned representation rather than the product-formula target
bias.

To check numerical stability at the very small success probabilities
in Table \ref{tab:sizes}, we also evaluate normalized evolution directly
under $\widetilde{H}$ with its lowest eigenvalue shifted to zero.
This shift leaves the normalized output unchanged and removes the
overall attenuation. The resulting mean infidelities reproduce the
tabulated values, showing that the large errors are not an artifact
of the small postselection normalization.

All tested XY configurations and the three-qubit XYZ configuration
have mean infidelities below 1\%. The four- and five-qubit XYZ configurations
remain above this level, including those with full diagonal support
and the largest tested depths. These results establish the performance
achieved with the chosen ansatz and finite optimization budget. In
particular, the five-qubit fit with $d=16$ reaches the 10,000-iteration
continuation limit. The remaining error therefore cannot be attributed
solely to ansatz expressivity, and these tests do not establish efficient
scaling to larger systems.

\begin{figure*}
\centering
\centering{}\includegraphics[width=1\textwidth]{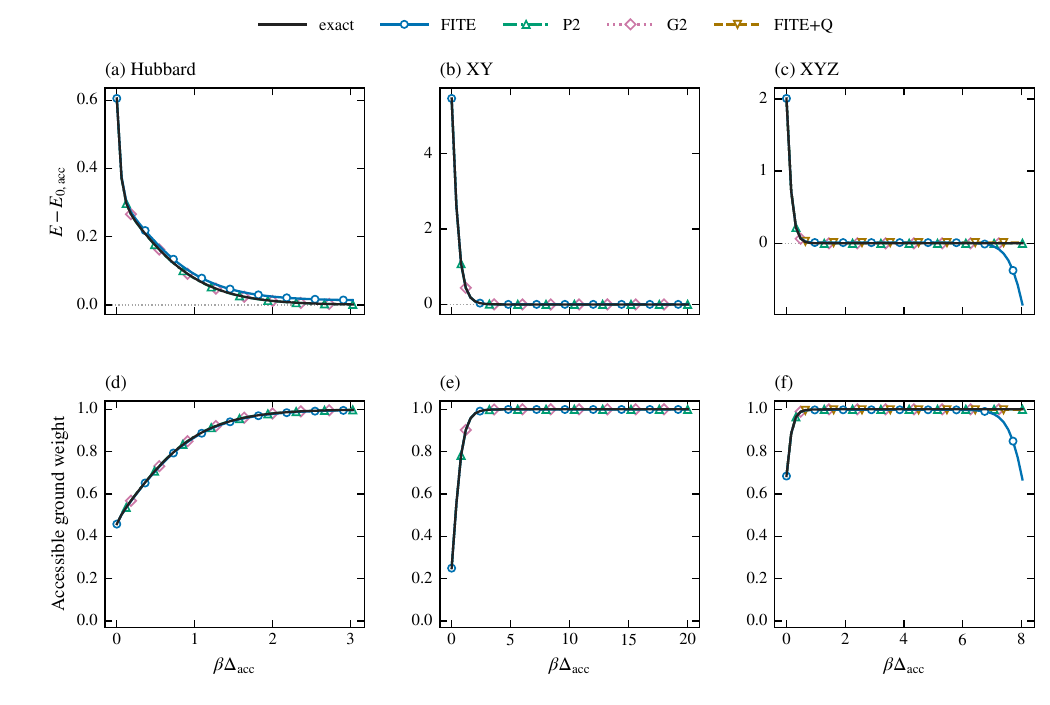}\caption{For Neel inputs, panels (a--c) show energy above the lowest initially
populated eigenspace, and panels (d--f) show its population. The
gap $\Delta_{{\rm acc}}$ separates this eigenspace from the next
populated one. The projector $Q$ restores the XYZ input parity; negative
energy excess indicates leakage.}\label{fig:physical}
\end{figure*}

\section{Physical observables and symmetry}\label{app:physical}

Figure \ref{fig:physical} and Table \ref{tab:observables} compare
the energy $E=\langle H\rangle$, bond-averaged correlation $\langle ZZ\rangle=(n-1)^{-1}\sum_{j=1}^{n-1}\langle Z_{j}Z_{j+1}\rangle$
and low-energy weight for $\ket +^{\otimes n}$ and Neel inputs, using
exact ITE of the same input as the reference. Here, $E_{0,{\rm acc}}$
is the lowest initially populated energy, $\Delta_{{\rm acc}}$ is
the gap to the next populated energy, and accessible weight is the
probability in the lowest populated eigenspace.

\begin{table}
\centering
\centering{}\caption{The table reports errors relative to exact ITE of the same input at
$\beta=5$. The quantities $\Delta E$ and $\Delta\langle ZZ\rangle$
are the energy and bond-averaged correlation differences, and $P=P(\psi)$
is the input-specific success probability. Hub. denotes Hubbard.}
\label{tab:observables}{\small{}%
\begin{tabular}{lrrrrr}
\hline 
Model & Input & Method & $|\Delta E|$ & $|\Delta\langle ZZ\rangle|$ & $P$\tabularnewline
\hline 
Hub. & plus & FITE & 0.013 & 0.0018 & 0.78\tabularnewline
Hub. & Neel & FITE & 0.013 & 0.0018 & 0.46\tabularnewline
XY & plus & FITE & 0.0017 & $3.2\times10^{-4}$ & 0.54\tabularnewline
XY & Neel & FITE & 0.0017 & $3.2\times10^{-4}$ & 0.25\tabularnewline
XYZ & plus & FITE & 0.01 & 0.0015 & $6.2\times10^{-16}$\tabularnewline
XYZ & Neel & FITE & 0.85 & $9.9\times10^{-5}$ & $2.7\times10^{-23}$\tabularnewline
XYZ & Neel & FITE+Q & 0.01 & 0.0015 & $1.8\times10^{-23}$\tabularnewline
\hline 
\end{tabular}}
\end{table}

For an input with parity $p_{\psi}=\pm1$, the projector $Q=(I+p_{\psi}\prod_{j=1}^{n}Z_{j})/2$
selects the same parity sector. FITE+Q applies this projector after
the FITE filter and accepts an output only when both postselections
succeed. Its Kraus operator is $Q\mathcal{K}_{{\rm FITE}}$, with
success probability $\|Q\mathcal{K}_{{\rm FITE}}\ket{\psi}\|_{2}^{2}$.

The XYZ input $\ket{010}$ has zero overlap with the global ground
state. At $\beta=5$, symmetry leakage gives an infidelity of 0.331.
Final parity projection reduces it to 0.0023, with success probability
$1.80\times10^{-23}$, at the cost of three additional CX gates and
one readout.

\selectlanguage{english}%
\bibliographystyle{apsrev4-2}
\bibliography{reference}
\selectlanguage{american}%

\end{document}